\documentstyle[aasms4]{article}

\slugcomment{{\em}}

\begin{document}

\title{On the Formation of Boxy and Disky Elliptical Galaxies}

\author{Thorsten Naab \& Andreas Burkert}
\affil{Max-Planck-Institut f\"ur Astronomie, K\"onigstuhl 17,\\
       D-69117 Heidelberg, \\Germany}
\author{Lars Hernquist}
\affil{Center for Astrophysics, 60 Garden Street,\\ Cambridge, MA 02138,\\
USA}

\begin{abstract}
The origin of boxy and disky elliptical galaxies is investigated. The
results of two collisionless N-body simulations of spiral-spiral
mergers with mass ratios of 1:1 and 3:1 are discussed and the 
projected properties of the merger remnants are investigated. It is
shown that the equal-mass merger leads to an anisotropic, slowly
rotating system with preferentially boxy isophotes and significant
minor axis rotation. The unequal-mass merger results in the formation 
of a rotationally supported elliptical with disky isophotes and small
minor axis rotation. The observed scatter in the kinematical and
isophotal properties of both classes of elliptical galaxies can be
explained by projection effects. 
\end{abstract}


\section{Introduction}
Elliptical galaxies have long been considered as old, coeval systems,
consisting of a dynamically relaxed, spheroidal stellar population 
with a universal $r^{1/4}$ surface brightness profile.
More detailed observations have however shown that these systems can
be subdivided into two classes with distinct kinematical
and orbital properties.
Low-luminosity ellipticals are isotropic and rotationally supported,
with small minor axis rotation and disky
deviations of their isophotal shapes from perfect ellipses
(\cite{b1988}; \cite{bdm1988}, hereafter BDM; \cite{kb1996} and
references therein). High-luminosity
ellipticals, on the other hand, are anisotropic, slow rotators with
large minor axis rotation, boxy isophotes and, occasionally, 
with kinematically distinct cores.
Bender el al. (1989)
demonstrated that both groups have different radio and X-ray
luminosities and recent high-resolution observations exhibit that disky ellipticals
have steep power-law cores in contrast to boxy ellipticals with flat cores
and central density cusps (\cite{l1995}, \cite{f1997}).

On the theoretical side, Toomre \& Toomre (1972) proposed that early type galaxies
originate from major mergers
of disk galaxies. This "merger hypothesis" has been investigated in great details
by numerous authors using numerical simulations (see \cite{bh1992}
for a review).  The first fully self-consistent merger models
of two equal-mass rotationally supported stellar disks, 
embedded in dark matter halos were performed by Barnes (1988) and  Hernquist (1992).
They found that mergers indeed 
lead to a slowly rotating, pressure supported, anisotropic spheroidal 
system. In the inner regions, the remnants were however too diffuse,
leading to strong deviations from the observed de Vaucouleurs profiles which
requires an inner $r^{-1}$ density profile. This result 
can be explained by phase space limitations (\cite{c1986}).
Subsequent investigations by Hernquist et al. (1993) showed that
mergers of progenitors with  massive  bulge components 
could resolve this problem, leading to ellipticals with small
core radii and surface brightness profiles that are in excellent agreement with observations.
Hernquist (1993b) and subsequently Heyl, Hernquist \& Spergel (1994) and Steinmetz
\& Buchner (1995) noted departures from pure ellipses in their equal-mass merger remnants. 
The same remnant when viewed from different
orientations appeared either boxy or disky. This seems to be in contradiction
with the observations. As disky and boxy ellipticals have different
radio and X-ray properties that should not depend on viewing angle their
isophotal shapes cannot change as a result of projection effects. In
agreement with observations  of boxy ellipticals, Barnes(1992) and
Heyl et al. (1996) also found misalignments  
between the spin and minor axis in major merger remnants which however seemed 
to be larger than observed (\cite{f1991}).

It has been argued by
Kormendy \& Bender (1996) and Faber et al. (1997) that gaseous mergers lead to distinct inner 
gaseous disks in the merger remnants which subsequently turn into stars, generating 
disky isophotes.  In contrast, boxy ellipticals would form from purely dissipationless 
mergers.  This idea has been addressed in details by
Bekki \& Shioya (1997) and Bekki (1998). Bekki \& Shioya (1997) simulated mergers including
gaseous dissipation and star formation. They found that the rapidity of gas consumption 
affects the isophotal shapes. Secular star formation however leads to final density
profiles which deviate significantly from the observed r$^{1/4}$-profiles in radial
regimes where all ellipticals show almost perfect de Vaucouleurs laws (Burkert 1993).
These calculations and models by Mihos \& Hernquist (1996) demonstrate
that the effect of  gas and star formation changes the 
structure of merger remnants as such a dissipative component would
most likely lead to strong deviations from the r$^{1/4}$-profiles which seems to be a result
of dissipationless, violent relaxation processes. Nevertheless the
observations of metal enhanced, decoupled and rapidly spinning
disk-like cores (\cite{bs1992}; \cite{dsp1993}; \cite{bd1996}) shows that even in boxy
ellipticals gas must have present. Numerical simulations show, that
these features would result naturally from gas infall during the
merger process (\cite{bh1996}; \cite{mh1996}). The influence of gas on the global structure of
elliptical galaxies is not well understood as it is sensitive to
uncertain details about star formation (\cite{bh1996}).   

Recently Barnes (1998) proposed a scenario of the origin of rapidly
rotating ellipticals that does not require gaseous dissipation. He
showed that such systems would result from the merger of a large disk
galaxy with a smaller companion. In addition, the edge on view shows a
disky morphology. Taking this model into account, boxy and
disky ellipticals should result from equal- and unequal-mass mergers,
respectively.  

In this letter the Barnes hypothesis is investigated in greater details.
We present the results of two merger simulations, a 1:1 merger and a 3:1 merger. 
The definition of the global orbital and kinematical properties depends strongly
on the method used. We therefore apply the same data reduction method to our model
galaxies which is used to derive the global parameters of observed ellipticals
and with which we compare our results.
The equal-mass merger indeed leads to a boxy elliptical whereas the
unequal-mass merger forms a disky elliptical, with detailed
kinematical properties that are in perfect agreement with the
observations.  

\section{The merger models}

The spiral galaxies are constructed in dynamical equilibrium 
using the method described by Hernquist (1993a).  
We use the following system of units: gravitational constant G=1, exponential
scale length of the larger disk h=1 and mass of the larger disk $M_d=1$.
Each galaxy consists of an exponential disk, a spherical, non-rotating
bulge with mass $M_b = 1/3$, a Hernquist density profile (\cite{her1990})
and a scale length $r_b=0.2h$  and a spherical pseudo-isothermal halo with a mass 
$M_d=5.8$, cut-off radius $r_c=10h$ and core radius $\gamma=1h$. 

The N-body simulations were performed using a direct summation code
with the special purpose hardware GRAPE (\cite{sug1990}). 
The 1:1 merger was calculated adopting in total 400000 particles with each galaxy
consisting of 20000 bulge
particles, 60000 disk particles and 120000 halo particles. For the 3:1
merger the parameters of the more massive galaxy were as described above. The low-mass
galaxy contained 1/3 the mass and number of particles in each component, with a disk 
scale length of $h=\sqrt{1/3}$, as expected from the Tully-Fisher relation.
For the  gravitational 
softening we used a  value of $\epsilon = 0.07$. 
In agreement with Walker et al. (1996) we noticed a growing
bar mode in the disk for test cases where we evolved galaxies in
isolation. This effect is however reduced
considerably with respect to previous calculations, due to our choice of twice
as many halo particles than disk particles.

For both mergers, the galaxies approach each other on nearly parabolic
orbits with an initial separation of 30 length units and a pericenter
distance of 2 length units. The inclinations of the
two disks relative to the orbit plane are $t_1= 30^{\circ}$ and $t_2
=-30^{\circ}$ with arguments of pericenter of $\omega_1 = 30^{\circ}$ and $ \omega_2
=-30^{\circ}$. These values are most likely for random encounters.
In both simulations the merger remnants were allowed to settle into equilibrium for
approximately  10 dynamical times after the merger was complete. Then their
equilibrium state was analysed.

\section{Analysis of the equilibrium states}
In order to compare the simulation with observations we 
follow as closely as possible the
analysis used by BDM. An artificial
image of the remnant is created by binning the central 10 length units into
$128 \times 128$ pixels. 
This picture is smoothed with a Gaussian filter of standard deviation 1.5
pixels. The isophotes and their deviations from perfect ellipses are then
determined using a reduction package kindly provided by Ralf Bender.  

Figure 1 shows the radial
distribution of $a_4$ along the major axis for 200 random projections.
 There is a clear trend for the 1:1 merger to have boxy (negative
$a_4$) isophotes while the 3:1 merger has disky
deviations (positive $a_4$) inside one half-mass radius.
Following the definition of BDM for the global
properties of observed elliptical galaxies, 
we determine for every projection $a_4$   as the
mean value between $0.25 r_e$ and $1.0 r_e$ , with $r_e$ being the
spherical half light radius. In case of a strong
peak in the $a_4$-distribution with an absolute value
that is larger than the absolute mean value, we choose the peak value.
In Figure 2 representative isophotal contours of the simulation of a
boxy (1:1 merger) and a disky (3:1 merger) merger remnant are shown.   

The characteristic ellipticity $\epsilon$ is defined as the isophotal ellipticity
at $1.5 r_e$. The central velocity dispersion $\sigma$ is determined as the
average projected velocity dispersion of the stars inside a projected
galactocentric distance of $ 0.2 r_e $. Finally, we define the characteristic rotational
velocity along the major and the minor axis as the projected rotational velocity
determined around $1.5 r_e$ and $0.5 r_e$, respectively. 

Figure 3 shows the characteristic isophotal and kinematical properties of the
1:1 merger (filled circles) and of the
3:1 merger (open circles) for 200 random projections. 
One can clearly see that the 1:1 merger produces remnants that are anisotropic
and boxy with large minor axis rotation whereas the 3:1 merger forms an isotropic and
disky elliptical with small minor axis rotation.
These results are in excellent agreement with the
observations of BDM.

\section{Discussion and Conclusions}

Our calculations demonstrate that the observed dichotomy between boxy and disky
ellipticals could originate from variations in the mass ratios of the merger
components. In general, the isophotal shapes change with radius.
An analysis applying methods and definitions
similar to those used for observed ellipticals does however lead to a
clear separation of properties between the models of equal and unequal-mass mergers in agreement with the observed structure 
of boxy and disky ellipticals, respectively. A trend for  equal-mass mergers to form
preferentially boxy ellipticals has already been noted before by
Steinmetz \& Buchner (1995), who did however neglect the bulge
component.
We find that a 3:1 merger is still efficient enough in order to disrupt the disks,
leading to spheroidal galaxies with de Vaucouleurs profiles. Our
analysis also shows that  these systems are  rotationally
supported with disky isophotes in the region inside one effective radius.

Projection effects do not change the fundamental difference between
equal and unequal-mass merger remnants. They do however lead to a
large spread in the global parameters. This is in very good agreement
with the observed parameter distribution (BDM; \cite{bsdm1989}). 

In contradiction to the common believe that disky E/S0 galaxies
are formed involving dissipative processes like star formation (\cite{bek1997})
we conclude that pure stellar mergers can in principle explain the observed 
dichotomy between disky and boxy ellipticals. This result is
supported by additional merger simulations with varying  mass ratios
and orbital parameters which will be discussed in more details in a subsequent paper. 

Observations show that disky ellipticals have on average lower
luminosities than boxy ellipticals (BDM). Our results would
indicate that low-mass elliptical galaxies preferentially formed
by unequal-mass mergers of disk-galaxies whereas
equal-mass mergers dominated the formation of high-mass
ellipticals. This result is puzzeling as there does not exist a
convincing argument for why low mass ellipticals should have suffered
mainly minor mergers while high mass ones should have evolved mainly
through major mergers. 

Since dissipative features are observed in all types of elliptical
galaxies in can not generally be ruled out, that the gas dynamics and
star formation have played an important role for the formation of boxy
and disky ellipticals.

\acknowledgments
 
We thank Ralf Bender, Hans-Walter Rix and Ralf Klessen for helpful disussions.

\clearpage

\begin{figure}
\plotone{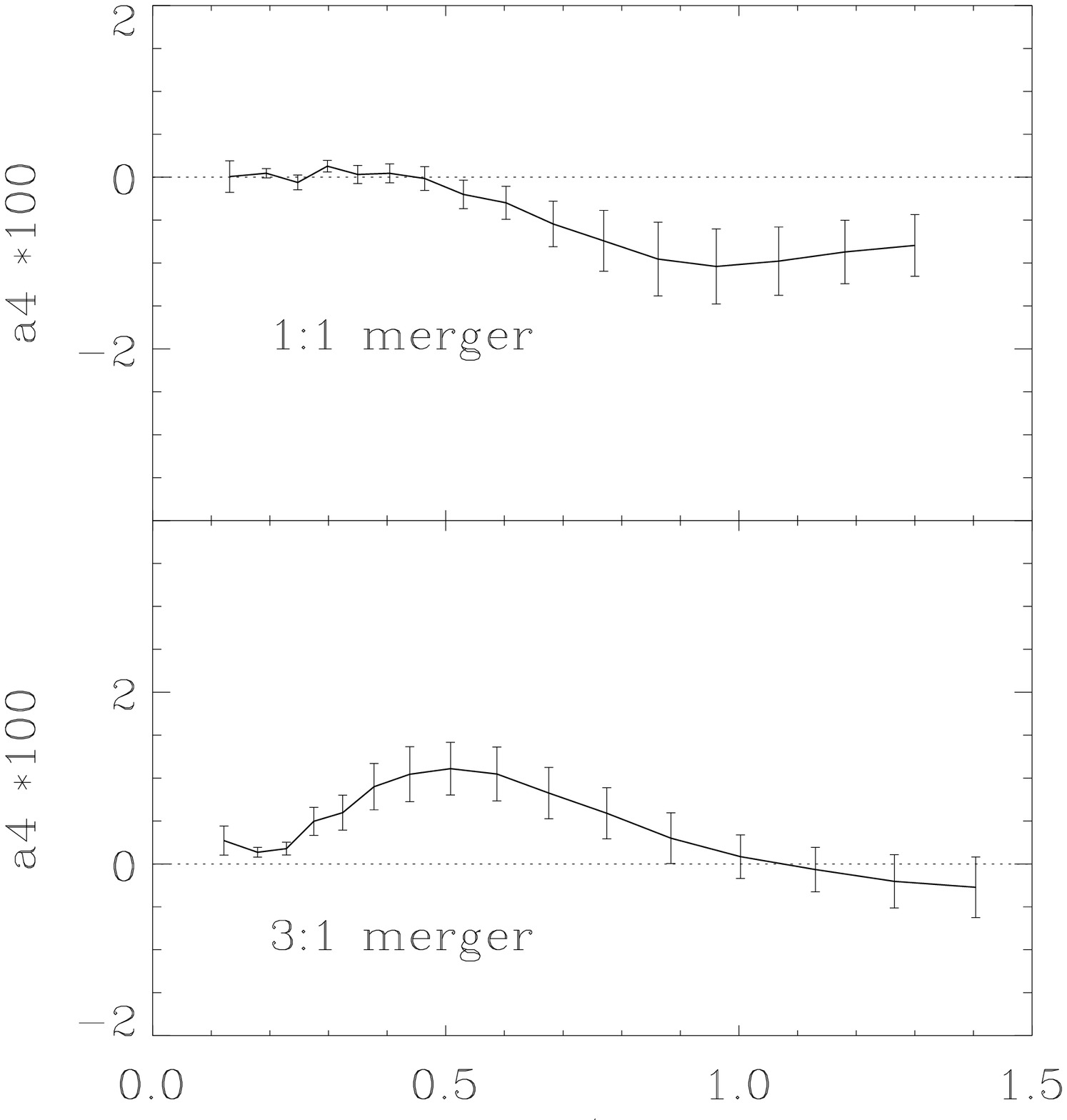} 
\caption
{Mean variation of $a_4$ along the apparent radius for 200 random
projections with $1 \sigma$ error bars which indicate the variation
due to projection effects. 
The 1:1 merger leads to a boxy isophotes (upper panel), while the 3:1
has positive values of $a_4$ (disky isophotes) inside one half light
radius $r_e$.\label{fig1}}
\end{figure}

\begin{figure}
\plotone{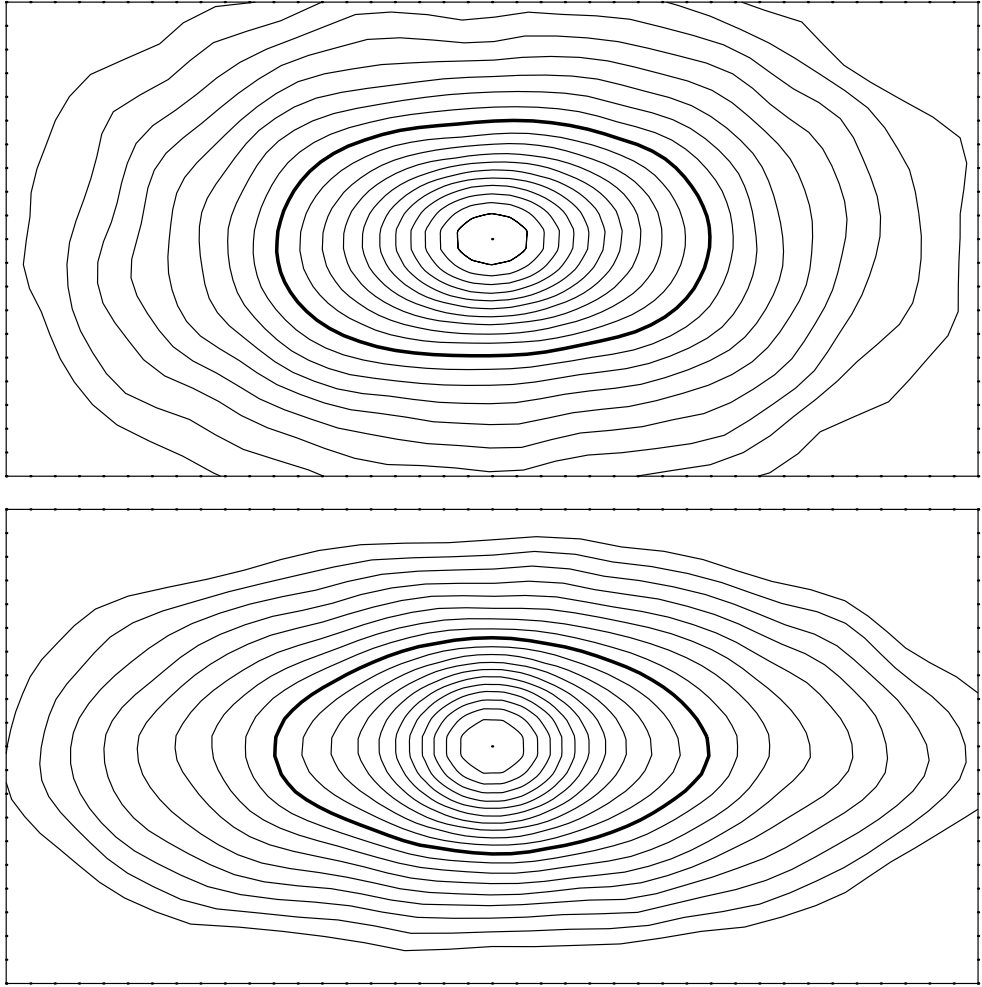} 
\caption
{Representative isophotal contours of a simulated 1:1 merger remnant
with boxy isophotes (upper panel) and a 3:1 merger remnant with disky 
isophotes (lower panel). The thick contour indicates the region where
$a4$ was determined.\label{fig2}}
\end{figure}

\begin{figure}
\plotone{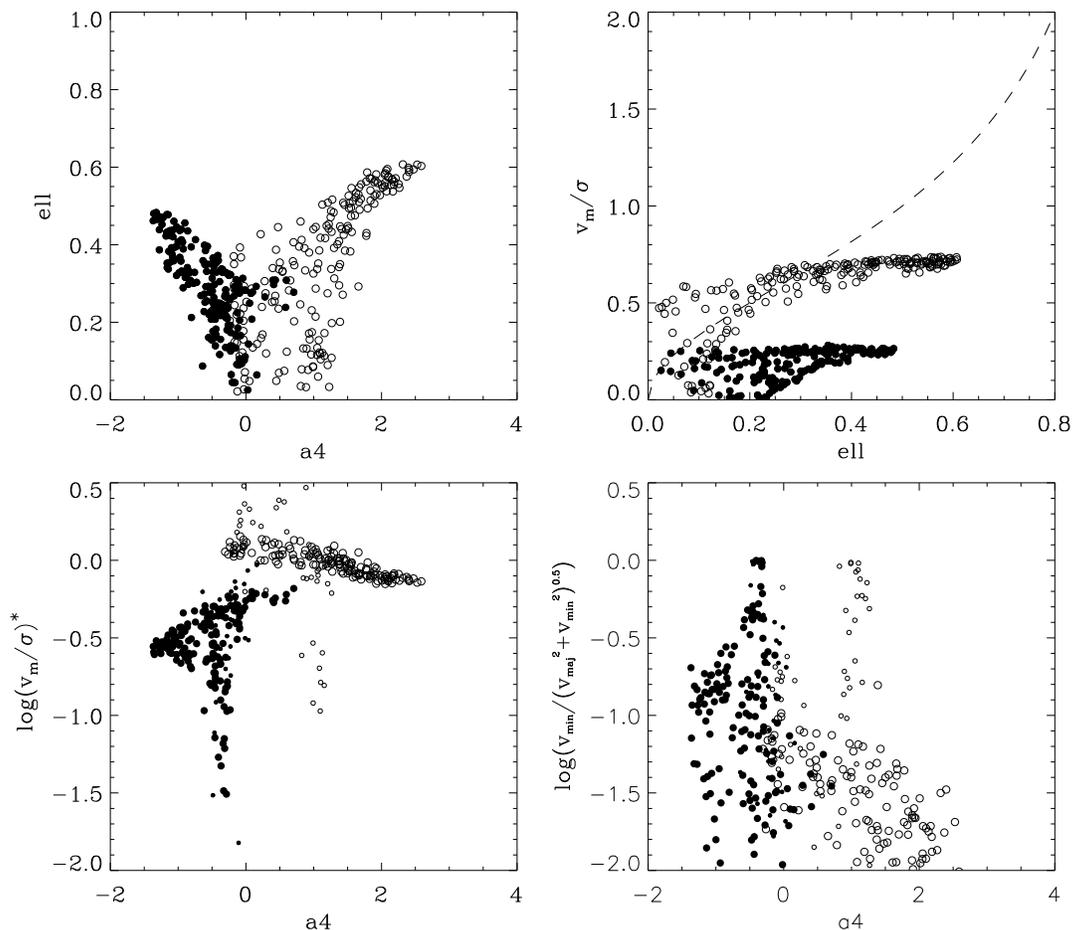} 
\figcaption	
{Kinematical and photometric properties for 200 random projections
of the merger remnants. Filled circles show the values for the equal-mass merger
while open circles show the values for the 3:1 merger. 
{\it Upper left panel}: Ellipticity of the remnants versus $a_4$. 
{\it Upper right panel}: Rotational velocity over central velocity
dispersion versus ellipticity. The dashed line shows the theoretical
curve for an oblate isotropic rotator.
{\it Lower left panel}: Anisotropy parameter $(v/\sigma)^*$ versus $a_4$.
The small dots and circles indicate projections with apparent
ellipticities smaller than $0.2$. For such values the determination of
the major axis is very uncertain, leading to large errors. 
{\it Lower left panel}: Amount of minor axis rotation versus $a_4$
with $v_{maj}$ an $v_{min}$ being the maximum velocity along the major
and minor axis, respectively.\label{fig3}}
 
\end{figure}
 
\end{document}